\documentclass{emulateapj}
\slugcomment{{\sc Accepted to ApJL:} June 19, 2012} 

\usepackage{amsmath}

\shorttitle{Asymmetric CSM in SNe Ia.}
\shortauthors{F\"orster et~al.}

\begin{document}

\title{\bf Evidence for asymmetric distribution of circumstellar
  material around Type Ia supernovae.}  \author{Francisco
  F\"orster$^1$, Santiago Gonz\'alez--Gait\'an$^1$, Joseph
  Anderson$^1$, Sebasti\'an Marchi$^1$, Claudia P. Guti\'errez$^1$, Mario
  Hamuy$^1$, Giuliano Pignata$^2$, R\'egis Cartier$^1$} \affil{$^1$
  Departamento de Astronom\'\i a, Universidad de Chile, Casilla 36-D,
  Santiago, Chile} \affil{$^2$Departamento Ciencias F\'isicas,
  Universidad Andres Bello, Av. Rep\'ublica 252, Santiago, Chile}

\begin{abstract}

\noindent We study the properties of low--velocity material in the
line of sight towards nearby Type Ia Supernovae (SNe Ia) that have
measured late phase nebular velocity shifts ($v_{\rm neb}$), thought
to be an environment--independent observable.  We have found that the
distribution of equivalent widths of narrow blended Na I D1 \& D2 and
Ca II H \& K absorption lines differs significantly between those SNe
Ia with negative and positive $v_{\rm neb}$, with generally stronger
absorption for SNe Ia with $v_{\rm neb} \ge 0$. A similar result had
been found previously for the distribution of colors of SNe Ia, which
was interpreted as a dependence of the temperature of the ejecta with
viewing angle. Our work suggests that: 1) a significant part of these
differences in color should be attributed to extinction, 2) this
extinction is caused by an asymmetric distribution of circumstellar
material (CSM) and 3) the CSM absorption is generally stronger on the
side of the ejecta opposite to where the ignition occurs.

Since it is difficult to explain 3) via any known physical processes
that occur \emph{before} explosion, we argue that the asymmetry of the
CSM is originated \emph{after} explosion by a stronger ionizing flux
on the side of the ejecta where ignition occurs, probably due to a
stronger shock breakout and/or more exposed radioactive material on
one side of the ejecta. This result has important implications for
both progenitor and explosion models.

\end{abstract}


\section{INTRODUCTION}

Type Ia Supernovae (SNe Ia) are important tools for understanding the
evolution of the Universe because of their high luminosities and
light--curve homogeneity, which led to their standardization for
cosmological distance measurements \citep{1993ApJ...413L.105P,
  1996AJ....112.2408H} and the discovery of the acceleration of the
Universe \citep{riess98, perl99}. They are also important because of
their complex nucleosynthetic output and high ejecta kinetic energies
($10^{51}$ erg), which make them key ingredients for galaxy evolution
theory.

Unfortunately, we still lack a clear understanding of the nature of SN
Ia progenitors. Although it is generally accepted that their
progenitors are carbon oxygen white dwarfs (CO WD) in mass transfering
binary systems \citep[for a review, see][]{2000ARA&A..38..191H}, there
is no agreement on what their companions are, whether the CO WD
reaches the Chandrasekhar mass at ignition, what the mechanism that
triggers the ignition is and how the CO WD burns to form the ejecta.

Perhaps the two scenarios that best match observed SN Ia spectra and
light curves are those considered in \citet{2012ApJ...750L..19R}:
either a stably accreting carbon oxygen white dwarf (CO WD) that
reaches the Chandrasekhar mass ($M_{\rm Ch}$), undergoes a
thermonuclear runaway and burns in a deflagration to detonation
transition, in the $M_{\rm Ch}$--single degenerate (SD) scenario
\citep{1984ApJ...286..644N, 1991A&A...245L..25K}; or a pair of CO WDs
that merge and ignite dynamically in the violent double degenerate
(DD) scenario \citep{2012ApJ...747L..10P}.

Two significant differences between both scenarios are the typical
central densities at ignition and the distribution of circumstellar
material (CSM) around their progenitor systems. In the $M_{\rm
  Ch}$--SD scenario the central density at ignition is sufficiently
high to avoid the synthesis of radioactive $^{56}$Ni in favor of
stable iron group elements (IGEs) in the innermost regions of the
ejecta. The $M_{\rm Ch}$--SD scenario has potentially abundant CSM
that could be produced by either the wind of the donor star, weak nova
explosions experienced by the accretor, or by mass loss due to
inefficient mass transfer. In the violent DD--scenario, conversely,
the primary central density is never high enough to avoid the
synthesis of radioactive $^{56}$Ni in the innermost regions of the
ejecta, and since the merger and ignition happen in a dynamical
time--scale there is no time to leave a significant imprint in the
CSM.

The $M_{\rm Ch}$--SD scenario appears to over--produce IGEs
\citep{1984ApJ...286..644N}, which are not produced significantly in
the violent DD scenario. Observations of flat--topped profiles of Fe
II lines in the optical and NIR at late phases has been argued to be
evidence for stable IGEs in the innermost regions of the ejecta
\citep{2004ApJ...617.1258H, 2007ApJ...661..995G}. Moreover,
significant nebular velocity shifts ($v_{\rm neb}$) measured in these
flat--topped profiles have been suggested as evidence for off--center
ignition scenarios, which is expected in $M_{\rm Ch}$--SD scenarios
\citep[see e.g.][]{2012ApJ...745...73N}. The relation between $v_{\rm
  neb}$ and other properties such as the rate of evolution of the
velocity of Si absorption lines (the \emph{velocity gradient}) or the
colors during the photospheric phase of evolution \citep{maeda10,
  2011MNRAS.413.3075M, 2011A&A...534L..15C} seems to lend support to
this picture.

For the nearby SN 2011fe, there is evidence for lack of interaction
between the ejecta and a companion star or a companion star wind. The
radius of the companion star has been constrained to be less than 0.1
$R_{\odot}$ \citep{2012ApJ...744L..17B, 2011Natur.480..344N,
  2011arXiv1110.2538B}, and its mass loss, less than $\approx 10^{-8}$
$M_{\odot}$ yr$^{-1}$ \citep{2012ApJ...746...21H}. However,
\citet{2011Sci...333..856S} showed that SNe Ia show an excess of
blue-shifted narrow Na lines, which is evidence for CSM around type Ia
progenitors ejected before explosion. Particular cases, specifically
SN 2002ic \citep{2003Natur.424..651H}, SN 2006X
\citep[][c.f. \citealt{2008AstL...34..389C}]{2007Sci...317..924P}, SN
1999cl \citep{2009ApJ...693..207B} and SN 2007le
\citep{2009ApJ...702.1157S} have also provided evidence for CSM around
SNe Ia, although in the case of SN 2002ic the nature of the explosion
is unclear \citep{2006ApJ...653L.129B}.

\citet{2009ApJ...702.1157S} noted that SNe Ia with variable narrow
absorption lines tend to have broad lines and high velocity
gradients. \citet{2012ApJ...748..127F} recently suggested that higher
velocity at peak, redder SNe Ia are preferentially found in lower mass
host galaxies, but also that they have an excess of blue-shifted
absorption systems \citep{2012arXiv1203.2916F}. These observations
highlight the importance of understanding the relation between
progenitor systems and dust properties for cosmological distance
determinations \citep[e.g.][]{2008ApJ...686L.103G,
  2010AJ....139..120F}. Since SNe Ia with $v_{\rm neb} \ge 0$ are
known to have higher velocity gradients, generally associated to
higher velocities at peak, and to be redder \citep{maeda10,
  2011MNRAS.413.3075M}, we investigate the relation between the
presence or absence of narrow Na I D1 \& D2 and Ca II H \& K
absorption lines (hereafter Na and Ca lines), a proxy for material in
the line of sight, with $v_{\rm neb}$, a geometrical proxy that should
depend only on viewing angle \citep{maeda10} and that should therefore
be independent of any evolutionary or host galaxy effects.

\section{SAMPLE SELECTION AND DATA ANALYSIS}

In this work we have used the sample of 25 nearby SNe Ia with measured
nebular velocity shifts ($v_{\rm neb}$) derived by
\citet{2012AJ....143..126B} and \citet{2011MNRAS.413.3075M} for which
we have measured equivalent widths (EWs) of Na (24 SNe) and Ca (23
SNe) absorption systems. The EWs were obtained from mid--resolution
spectra in the public Center for Astrophysics (CfA) spectra database
\citep{2012AJ....143..126B, 1999ApJS..125...73J, 2003AJ....126.1489B,
  2008AJ....135.1598M, 2009ApJ...697..380W} for most SNe Ia, from
\citet{2011MNRAS.410..585S} for SN2009dc, from high--resolution
spectra in the European Southern Observatory (ESO) archive for SN
2000cx, SN 2001el and SN 2006X, from \citet{2011Sci...333..856S} for
SN 2007sr and SN 2007af, using \citet{2012arXiv1204.1891Y}, values
from the literature for SN 1986G, SN 2004eo and SN 2006dd
\citep{1993ApJ...413L.105P, 2007MNRAS.377.1531P, 2010AJ....140.2036S},
and our own X--Shooter \citep{2011A&A...536A.105V} data for SN
2010ev. We note that although there are public spectra for SN 1990N
and SN 2004eo, we do not use them because they appear to have been
artificially smoothed in the wavelength direction, removing all
evidence for narrow lines.

\subsection{Equivalent width determination}

We have written a Python script to measure EWs in every SN for every
epoch available. The script first defines the continuum by tracing
straight lines between nodes automatically defined. The position of
the nodes is chosen to avoid being too close to the lines being
measured, and the flux associated to every node is defined by locally
averaging the observed flux with typical velocity ranges proportional
to the separation between nodes and a continuous smoothing
function. After the nodes are placed and the continuum defined, we
visually inspect the spectra, divide the observed flux by the
continuum and compute the area below unity within 300 km s$^{-1}$ of
every line in $\rm \AA$:
\begin{equation} \label{eq:ew}
  {\rm EW} = \sum_{i} \frac{\bar f_i - f_i}{\bar f_i} ~d\lambda = \sum_{i}
  \bigg( 1 - \frac{f_i}{\bar f_i} \bigg) ~d\lambda,
\end{equation}
where $f_{i}$ are the fluxes of pixels with velocities within 300 km
s$^{-1}$ of the line of interest, $\bar f_i$ is the continuum value
for every pixel, and $d\lambda$ is the wavelength separation between
pixels. In order to obtain more robust EW determinations, for every
SN we use the median EW among all available epochs, excluding
those epochs with significantly larger errors.

We have considered four sources of error in our method somewhat
conservatively: flux measurement uncertainties ($\sigma_{f}$),
continuum flux uncertainties ($\sigma_{c}$), possible systematic
deviations in our method ($\sigma_{\rm sys}$), and the median absolute
deviation (a robust error measurement) of the EWs for different epochs
for every SN ($\sigma_{\rm MAD}$). $\sigma_{f}$ and $\sigma_{c}$ were
obtained using the quoted flux uncertainties for every spectra at
different wavelengths and propagating errors using
equation~(\ref{eq:ew}) to compute the EW uncertainty, $\sigma_{\rm
  EW}$:
\begin{equation} \label{eq:dew}
  \sigma_{\rm EW}^2 = \sum_{i} \bigg[ \bigg( \frac{\sigma_{f_i}}{\bar f_i} \bigg)^2 +
  \bigg(\frac{f_i}{\bar f_i} \frac{\sigma_{\bar f_i}}{\bar f_i} \bigg)^2 \bigg] ~d\lambda^2 \equiv
  \sigma_f^2 + \sigma_c^2,
\end{equation}
where $\sigma_{f_i}$ are the flux uncertainties in every pixel,
$\sigma_{\bar f_i}$ is the uncertainty associated to the continuum at every
pixel, defined as the average node flux uncertainty of the closest two
nodes to every pixel, and where each node uncertainty is defined by
weighting the $\sigma_{f_i}$'s in the same way as the flux of every
node is defined.

To determine $\sigma_{\rm sys}$ we have run our script assuming that the
lines are in positions of the spectra where no low velocity absorption
lines should be found, changing these positions artificially along the
spectra and using quadratically averaged values of the derived EWs as
the final errors. To determine $\sigma_{\rm MAD}$ we simply computed
the median absolute deviation of the measured EWs at different epochs,
excluding those values with relatively large errors. 

\subsection{Validation of the method}

\begin{figure}[h!]
\epsscale{1.25}
\plotone{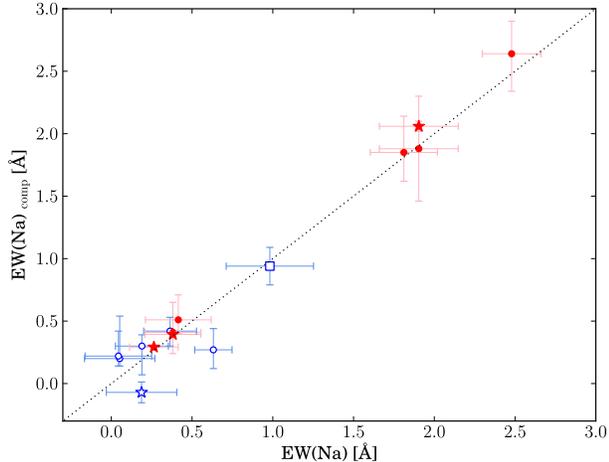}
\caption{Comparison of measured equivalent widths of blended Na
  absorption features due to intervening material at the distance of
  the SNe Ia. We show the values obtained with our Python script using
  mid--resolution spectra, EW(Na), and compare them with those in
  \citet{2009ApJ...693..207B}, \emph{circles},
  \citet{2011MNRAS.412.2735T}, \emph{square}, and high--resolution
  spectra measurements, \emph{stars}. Red filled symbols are SNe Ia
  with $v_{\rm neb} \ge 0$ and blue open symbols, SNe Ia with $v_{\rm
    neb} < 0$. The dotted line shows the EW(Na)$_{\rm comp}$ = EW(Na)
  relation. The only SN Ia with significantly different measurements
  in our sample is SN 1998bu, with $v_{\rm neb} < 0$, and for which we
  measure a higher EW than \citet{2009ApJ...693..207B}.}
\label{fig:comp}
\end{figure}

\begin{deluxetable} {lrrr}
\tablecolumns{4} \tablenum{1} \tablewidth{0pc} \tablecaption{Supernovae
  with available nebular velocities and their measured equivalent
  widths (EWs) for Na and Ca lines in \AA ~as described in the
  text. NA indicates that data is not available. }
\tablehead{ \colhead{Name} & \colhead{$v_{\rm neb}$ [km s$^{-1}$]} & \colhead{EW(Na)} & \colhead{EW(Ca)}}
\startdata
SN 1986G  &  -639 $\pm$ 600 &  3.60 $\pm$ 0.01 &  1.75 $\pm$ 0.01 \\
SN 1990N  &  -478 $\pm$ 600 &  NA              &  NA              \\
SN 1994D  & -2220 $\pm$ 220 &  0.05 $\pm$ 0.22 &  0.06 $\pm$ 0.25 \\
SN 1994ae & -1204 $\pm$ 600 &  0.36 $\pm$ 0.16 &  0.33 $\pm$ 0.22 \\
SN 1995D  & -1193 $\pm$ 600 &  0.32 $\pm$ 0.21 &  0.23 $\pm$ 0.29 \\
SN 1997bp &  2539 $\pm$ 410 &  1.81 $\pm$ 0.21 &  1.12 $\pm$ 0.26 \\
SN 1998aq & -1157 $\pm$ 129 &  0.19 $\pm$ 0.16 &  0.35 $\pm$ 0.20 \\
SN 1998bu & -1351 $\pm$ 305 &  0.63 $\pm$ 0.12 &  0.07 $\pm$ 0.16 \\
SN 2000cx &  -172 $\pm$ 600 & -0.07 $\pm$ 0.08 & -0.01 $\pm$ 0.01 \\
SN 2001el &   980 $\pm$ 206 &  0.66 $\pm$ 0.04 &  0.67 $\pm$ 0.10 \\
SN 2002bo &  2258 $\pm$ 361 &  2.48 $\pm$ 0.18 &  1.20 $\pm$ 0.26 \\
SN 2002dj &  2021 $\pm$ 414 &  1.09 $\pm$ 0.33 &  0.78 $\pm$ 0.31 \\
SN 2002er &   582 $\pm$ 600 &  1.77 $\pm$ 0.17 &  1.62 $\pm$ 0.31 \\
SN 2003du &  -443 $\pm$ 283 &  0.07 $\pm$ 0.21 & -0.02 $\pm$ 0.22 \\
SN 2003hv & -2825 $\pm$ 384 & -0.09 $\pm$ 0.25 & -0.41 $\pm$ 0.41 \\
SN 2003kf &  1070 $\pm$ 201 &  0.41 $\pm$ 0.20 &  0.69 $\pm$ 0.25 \\
SN 2004dt & -2623 $\pm$ 600 &  0.23 $\pm$ 0.23 & -0.02 $\pm$ 0.36 \\
SN 2004eo &  -170 $\pm$ ~25 &  0.00 $\pm$ 0.20 &  NA              \\
SN 2005cf &  -123 $\pm$ 600 &  0.04 $\pm$ 0.21 &  0.23 $\pm$ 0.20 \\
SN 2006X  &  2832 $\pm$ 343 &  2.06 $\pm$ 0.04 &  0.80 $\pm$ 0.06 \\
SN 2006dd & -1553 $\pm$ 201 &  3.22 $\pm$ 0.08 &  1.00 $\pm$ 0.10 \\
SN 2007af &   831 $\pm$ 350 &  0.39 $\pm$ 0.02 &  0.28 $\pm$ 0.17 \\
SN 2007sr &  1185 $\pm$ 600 &  0.29 $\pm$ 0.01 &  0.09 $\pm$ 0.26 \\
SN 2009dc & -1190 $\pm$ 600 &  0.94 $\pm$ 0.15 &  2.00 $\pm$ 0.31 \\
SN 2010ev &  2145 $\pm$ 225 &  0.67 $\pm$ 0.05 &  0.37 $\pm$ 0.02 \\
\enddata
\end{deluxetable}

We have compared our measurements of the EWs with those in the
literature and with those obtained from publicly available
high--resolution spectra, as shown in Figure~\ref{fig:comp}. Our
results are consistent within errors in 14 out of 15 measurements,
with the only exception being SN 1998bu, for which we obtained a
significantly higher EW, that upon visual inspection appears to be due
to continuum definition uncertainties. We have also compared our
measurements with those obtained with IRAF's \emph{splot} routine and
found consistent values in 90\% of the cases, without any highly
significant exception. The nebular velocities and equivalent widths
used in this analysis are shown in Table~1.

\section{RESULTS}

\begin{deluxetable*} {rcccccc}
\tablecolumns{7} \tablenum{2} \tablewidth{0pc} \tablecaption{KS test
  for Na and Ca lines. The null hypothesis is that SNe Ia with
  negative and positive $v_{\rm neb}$ have the same distribution of
  colors or EWs for a given line due to the Milky Way (MW) or
  intervening material at the distance of the SNe.}  \tablehead{
  \colhead{KS test} & \colhead{E(B-V)$^{\rm SNooPy}$} &
  \colhead{(B-V)$_{\rm B_{\max}}^{\rm SiFTO}$} & \colhead{MW Na} &
  \colhead{Host Na} & \colhead{MW Ca} & \colhead{Host Ca}} \startdata
$p$--value & 0.006 & 0.002 & 0.268 & 0.013 & 0.402 & 0.030 \enddata
\end{deluxetable*}

In order to test whether SNe Ia with negative and positive $v_{\rm
  neb}$ have different distributions of intervening material in the
line of sight, we have performed Kolmogorov--Smirnov (KS) tests of the
cumulative distribution function (CDF) of the measured equivalent
widths of Na and Ca lines caused by intervening material in the Milky
Way and in the SN host galaxies. We have also tested whether SNe Ia
with negative or positive $v_{\rm neb}$ have different extinction or
color properties using E(B-V) values obtained with SNooPy
\citep{2011AJ....141...19B} and (B-V)$_{\rm B_{\max}}$ values from
SiFTO \citep{2008ApJ...681..482C} best--fitting models. The KS test
results are shown in Table~2.

\begin{deluxetable*} {rcccccc}
\tablecolumns{5} \tablenum{3} \tablewidth{0pc} \tablecaption{KS tests
  for environmental differences between SNe Ia with negative and
  positive $v_{\rm neb}$. The null hypothesis is that the distribution
  of global (morphological type and inclination) and local (NCR$_{\rm
    NUV}$ and NCR$_{\rm H\alpha}$) host properties distributions are
  the same.} \tablehead{ \colhead{KS test} & \colhead{Morphological
    type} & \colhead{Galaxy inclination} & \colhead{NCR$_{\rm NUV}$} &
  \colhead{NCR$_{\rm H\alpha}$} }\startdata $p$--value & 0.133 & 0.390
& 0.768 & 0.944 \\ \enddata
\end{deluxetable*}

As expected, the distribution of intervening Milky Way material
between SNe Ia with negative and positive $v_{\rm neb}$ is consistent
with being drawn from the same parent population. However, the
likelihood that the distribution of intervening material in their host
galaxies is the same is significantly smaller. Given that the
separation of the sample by the sign of $v_{\rm neb}$ should be purely
geometrical, we should not expect any significant differences in the
distributions of host Na and Ca lines, unless the intervening material is of
circumstellar nature and with an asymmetric distribution.

To test whether the different distributions of EWs are caused by an
unlikely combination of host galaxy properties, we compare the
distributions of host galaxy morphological types and inclinations from
the Asiago catalogue \citep{1999A&AS..139..531B} and the higher
resolution pixel based statistic NCR \citep{2006A&A...453...57J} that
quantifies any association with regions of star formation (SF) using
\emph{GALEX} NUV (23 out of 25 SNe) and our own $H\alpha$ maps (14 out
of 25 SNe), separating between SNe Ia with $v_{\rm neb} \ge 0$ and
$v_{\rm neb} < 0$. We do not find any significant difference between
these two populations for global (morphological type and inclination)
or local (NCR statistics) host galaxy properties. The KS test
$p$--values are shown in Table~3.

\begin{figure}[h!]
\epsscale{1.25}
\plotone{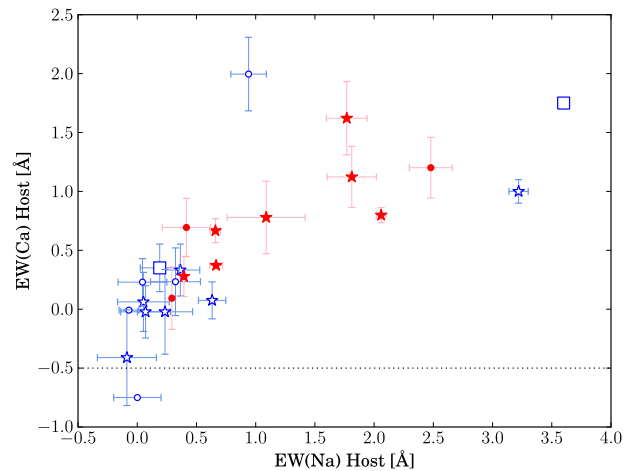}
\caption{Equivalent widths of Ca vs Na lines due to intervening
  material at the redshift of the SN host galaxies. We show SNe Ia
  with $v_{\rm neb} \ge 0$ as red filled symbols, and those with
  $v_{\rm neb} < 0$ as blue open symbols. SNe Ia occurring in pixels
  with NCR$_{\rm NUV} > 0$ are shown as stars, those with NCR$_{\rm
    NUV} = 0$ are shown as circles and those with no measured
  NCR$_{\rm NUV}$ are shown as squares. The three SNe Ia with $v_{\rm
    neb} < 0$ at high EWs are SN 1986G (open blue square), SN 2006dd
  (open blue star) and SN 2009dc (open blue circle). SN 2004eo, for
  which we only have EW(Na) information, is shown below the dotted
  line.}
\label{fig:ewHNa_ewHCa}
\end{figure}

In Figure~\ref{fig:ewHNa_ewHCa} we plot the measured EWs for the Na
and Ca lines divided by the sign of $v_{\rm neb}$. Red filled symbols
are SNe with $v_{\rm neb} \ge 0$ and blue open symbols, those with
$v_{\rm neb} < 0$. Stars indicate SNe Ia occurring in pixels
associated to SF (NCR$_{\rm NUV} > 0$), circles indicate SNe Ia
occurring in pixels with no association to SF (NCR$_{\rm NUV} = 0$)
and squares indicate SNe Ia with no NCR$_{\rm NUV}$ measurements. SN
2004eo, which only has a reported non--detection of Na in
\citep{2007MNRAS.377.1531P} and for which we associate a typical error
of 0.2 \AA, is shown below the dotted line.  In this figure, SNe Ia
with $v_{\rm neb} < 0$ appear clustered at relatively low EW values,
with three exceptions at high EWs, whereas $v_{\rm neb} \ge 0$ SNe Ia
tend to have higher EWs in general.  This suggests that the
intervening CSM absorbs more light opposite to the direction of where
the ignition occurs.

The fact that some SNe Ia show high EWs and $v_{\rm neb} < 0$ (3 out
of 14) is expected, since in some cases the EWs will be due to host
galaxy material and not CSM absorption. Even if their high EWs are due
to CSM our KS tests rule out that these exceptions are significant. In
particular, SN 1986G and SN 2006dd appear to be located behind
prominent dust lanes of their host galaxies and SN 2006dd has the
second highest NCR$_{\rm NUV}$ statistic in our sample (0.827),
implying a possible association with SF. However, SN 2006dd and SN
2009dc have been suggested to show some signs of CSM based on their Na
line evolution \citep{2010AJ....140.2036S} and light curve shape
\citep{2011MNRAS.412.2735T}, hence we cannot rule out that their high
EWs are due to CSM.

Most SNe Ia with $v_{\rm neb} \ge 0$ in Figure~\ref{fig:ewHNa_ewHCa}
have some degree of association with SF (NCR$_{\rm NUV} > 0$), which
could be associated to higher EWs due to non--CSM in the line of
sight. However, only 2 out of 10 SNe Ia with $v_{\rm neb} \ge 0$ are
strongly associated with SF (NCR$_{\rm NUV} \ge 0.5$), whereas 4 out
of 15 SNe with $v_{\rm neb} < 0$ are strongly associated to SF. This
suggests that higher EWs in our sample are not primarily due to host
galaxy non--CSM, which is consistent with the excess of blue--shifted
Na lines found by \citet{2011Sci...333..856S}, and which is supported
by the high $p$--values for the NCR$_{\rm NUV}$ and NCR$_{\rm
  H\alpha}$ KS tests in Table~3. It could also be argued that NCR
statistics are only a measure of relative SF within galaxies. To
account for this, we have removed early--type galaxies from our sample
and repeated the KS tests from Table~2, obtaining significantly low
$p$--values of 0.022 and 0.018 for the distribution of EW(Na) and
EW(Ca), respectively.

\section{DISCUSSION AND CONCLUSIONS}

Our main conclusion is that the distribution of EWs from narrow
blended Na and Ca lines differs significantly between SNe Ia with
negative and positive $v_{\rm neb}$. Because the sign of $v_{\rm neb}$
should be purely geometrical, it should not correlate with any host
galaxy or average SN properties.  This suggests that part of the lines
are formed by CSM ejected by the progenitor before explosion, which is
found asymmetrically distributed after explosion. The differences in
color found for SNe Ia with different $v_{\rm neb}$
\citep{2011MNRAS.413.3075M}, which we confirm using SNooPy and SiFTO
light curve fitting, should be explained in part by extinction due to
an asymmetric distribution of CSM, and not by different ejecta
temperatures with viewing angle alone.

There are two possibilities to explain this result: 1) the CSM
material is asymmetrically ejected from the system \emph{before}
explosion and is aligned with the side of the ejecta where ignition
occurs, or 2) the CSM material is initially spherically symmetric and
is affected by an asymmetric distortion \emph{after} explosion, which
is stronger on the side where ignition occurs.

One possible physical process that could break the symmetry of the WD
and have an effect on both the direction of ignition and the
distribution of CSM is rotation. However, rotation should produce
cylindrically symmetric systems, where positive or negative $v_{\rm
  neb}$ should lead to the same average CSM properties in the line of
sight. Another possibility is the gravitational field of a companion
star, but the typical velocity offsets found for the absorbing
material in \citet{2011Sci...333..856S} implies that the material was
ejected at a time before explosion much longer than the expected
orbital period of the progenitor systems, again producing some
cylindrical symmetry in average.

Thus, we conclude that only 2) is possible to explain our results. One
way to produce an asymmetric distribution of dust, EW(Na) and EW(Ca)
is to have an asymmetric ionizing field \emph{after} explosion. This
would destroy most of the dust and ionize Na and Ca in the direction
where ignition occurs. A comparison with pre--explosion images of the
core--collapse SN 2012aw \citep{2012arXiv1204.1523F} supports the idea
of dramatic changes in extinction after explosion. The fact that the
correlations are found to be stronger for dust (colors) than for
EW(Na) and EW(Ca) is suggestive, since they are in ascending order of
ionization potential. One problem with this interpretation could be
the constraints in radio and X--ray for CSM--ejecta interactions. This
would be solved if the material were placed far enough from the
ejecta, and only in some cases allow the ejecta to interact with the
CSM, as was seen in SN 2002ic, SN 2006X and 2007le, or if the measured
EWs are due to saturated lines in low mass, clumpy CSM with a wide
distribution of velocities.

Our results provide hints about the explosion asymmetry and CSM
properties of SN Ia progenitors. If a larger sample confirms that most
SNe Ia with $v_{\rm neb} \ge 0$ have high EWs of Na and Ca absorption
lines, the majority of progenitors should contain significant CSM. If
this is the case, most SNe Ia with $v_{\rm neb} < 0$ should have more
strongly ionizing fields (higher energy photons) after explosion,
either during shock breakout, or afterwards if more radioactive
material is exposed in the ignition side of the explosion.

Finally, significant CSM is consistent with some $M_{\rm Ch}$--SD or
slow DD scenarios \citep[e.g.][]{2010ApJ...725..296F} and not with
violent DD scenarios. However, in our interpretation it is difficult
to explain the SNe Ia with $v_{\rm neb} \ge 0$ and low EWs, which may
suggest that multiple progenitor scenarios are at work.

\acknowledgments

\noindent

We thank the referee for providing important feedback for this
work. F.F., J.A. and G.P acknowledge support from FONDECYT through
grants 3110042, 3110142 and 11090421. J.A., R.C., F.F, S.G., M.H. and
G.P. acknowledge support provided by the Millennium Center for
Supernova Science through grant P10-064-F (funded by ``Programa
Bicentenario de Ciencia y Tecnolog\'ia de CONICYT'' and ``Programa
Iniciativa Cient\'ifica Milenio de MIDEPLAN''). This research has made
use of the CfA Supernova Archive, which is funded in part by the
National Science Foundation through grant AST 0907903. We have used
The Weizmann interactive supernova data repository -
\verb+www.weizmann.ac.il/astrophysics/wiserep+.  Based on observations
made with the European Southern Observatory telescopes obtained from
the ESO/ST-ECF Science Archive Facility. Based on observations made
with the NASA Galaxy Evolution Explorer. GALEX is operated for NASA by
the California Institute of Technology under NASA contract NAS5-98034.


\begin{thebibliography}{}

\bibitem[Barbon et 
al.(1999)]{1999A&AS..139..531B} Barbon, R., Buond{\'{\i}}, V., Cappellaro, E., \& Turatto, M.\ 1999, \aaps, 139, 531 

\bibitem[Benetti et al.(2006)]{2006ApJ...653L.129B} Benetti, S., 
Cappellaro, E., Turatto, M., et al.\ 2006, \apjl, 653, L129 

\bibitem[Blondin et al.(2009)]{2009ApJ...693..207B} Blondin, S., Prieto, 
J.~L., Patat, F., et al.\ 2009, \apj, 693, 207 

\bibitem[Blondin et al.(2012)]{2012AJ....143..126B} Blondin, S., Matheson, 
T., Kirshner, R.~P., et al.\ 2012, \aj, 143, 126 

\bibitem[Bloom et al.(2012)]{2012ApJ...744L..17B} Bloom, J.~S., Kasen, D., 
Shen, K.~J., et al.\ 2012, \apjl, 744, L17 

\bibitem[Branch et al.(2003)]{2003AJ....126.1489B} Branch, D., Garnavich, 
P., Matheson, T., et al.\ 2003, \aj, 126, 1489 

\bibitem[Brown et al.(2011)]{2011arXiv1110.2538B} Brown, P.~J., Dawson, 
K.~S., de Pasquale, M., et al.\ 2011, arXiv:1110.2538 

\bibitem[Burns et al.(2011)]{2011AJ....141...19B} Burns, C.~R., 
Stritzinger, M., Phillips, M.~M., et al.\ 2011, \aj, 141, 19 

\bibitem[Cartier et 
al.(2011)]{2011A&A...534L..15C} Cartier, R., F{\"o}rster, F., Coppi, P., et al.\ 2011, \aap, 534, L15 

\bibitem[Chugai(2008)]{2008AstL...34..389C} Chugai, N.~N.\ 2008, Astronomy 
Letters, 34, 389 

\bibitem[Conley et al.(2008)]{2008ApJ...681..482C} Conley, A., Sullivan, 
M., Hsiao, E.~Y., et al.\ 2008, \apj, 681, 482 

\bibitem[Folatelli et al.(2010)]{2010AJ....139..120F} Folatelli, G., 
Phillips, M.~M., Burns, C.~R., et al.\ 2010, \aj, 139, 120 

\bibitem[Foley(2012a)]{2012ApJ...748..127F} Foley, R.~J.\ 2012, \apj, 748, 
127 

\bibitem[Foley et al.(2012b)]{2012arXiv1203.2916F} Foley, R.~J., Simon, 
J.~D., Burns, C.~R., et al.\ 2012, arXiv:1203.2916, \emph{Accepted for ApJ.}

\bibitem[Fraser et al.(2012)]{2012arXiv1204.1523F} Fraser, M., Maund, 
J.~R., Smartt, S.~J., et al.\ 2012, arXiv:1204.1523 

\bibitem[Fryer et al.(2010)]{2010ApJ...725..296F} Fryer, C.~L., Ruiter, 
A.~J., Belczynski, K., et al.\ 2010, \apj, 725, 296 

\bibitem[Gerardy et al.(2007)]{2007ApJ...661..995G} Gerardy, C.~L., Meikle, 
W.~P.~S., Kotak, R., et al.\ 2007, \apj, 661, 995 

\bibitem[Goobar(2008)]{2008ApJ...686L.103G} Goobar, A.\ 2008, \apjl, 686, 
L103 

\bibitem[Hamuy et al.(1996)]{1996AJ....112.2408H} Hamuy, M., Phillips, 
M.~M., Suntzeff, N.~B., et al.\ 1996, \aj, 112, 2408 

\bibitem[Hamuy et al.(2003)]{2003Natur.424..651H} Hamuy, M., Phillips, 
M.~M., Suntzeff, N.~B., et al.\ 2003, \nat, 424, 651 

\bibitem[Hillebrandt 
\& Niemeyer(2000)]{2000ARA&A..38..191H} Hillebrandt, W., \& Niemeyer, J.~C.\ 2000, \araa, 38, 191 

\bibitem[H{\"o}flich et al.(2004)]{2004ApJ...617.1258H} H{\"o}flich, P., 
Gerardy, C.~L., Nomoto, K., et al.\ 2004, \apj, 617, 1258 

\bibitem[Horesh et al.(2012)]{2012ApJ...746...21H} Horesh, A., Kulkarni, 
S.~R., Fox, D.~B., et al.\ 2012, \apj, 746, 21 

\bibitem[James \& Anderson(2006)]{2006A&A...453...57J} James, P.~A.,
  \& Anderson, J.~P.\ 2006, \aap, 453, 57

\bibitem[Jha et al.(1999)]{1999ApJS..125...73J} Jha, S., Garnavich, P.~M., 
Kirshner, R.~P., et al.\ 1999, \apjs, 125, 73 

\bibitem[Khokhlov(1991)]{1991A&A...245L..25K} Khokhlov, A.~M.\ 1991, \aap, 245, L25 

\bibitem[Kotak et al.(2004)]{2004MNRAS.354L..13K} Kotak, R., Meikle, 
W.~P.~S., Adamson, A., \& Leggett, S.~K.\ 2004, \mnras, 354, L13 

\bibitem[Maeda et~al.(2010)]{maeda10} Maeda, K. et~al. 2010, Nature, 486, 82

\bibitem[Maeda et al.(2011)]{2011MNRAS.413.3075M} Maeda, K., Leloudas, G., 
Taubenberger, S., et al.\ 2011, \mnras, 413, 3075 

\bibitem[Matheson et al.(2008)]{2008AJ....135.1598M} Matheson, T., 
Kirshner, R.~P., Challis, P., et al.\ 2008, \aj, 135, 1598 

\bibitem[Nomoto et al.(1984)]{1984ApJ...286..644N} Nomoto, K., Thielemann, 
F.-K., \& Yokoi, K.\ 1984, \apj, 286, 644 

\bibitem[Nonaka et al.(2012)]{2012ApJ...745...73N} Nonaka, A., Aspden, 
A.~J., Zingale, M., et al.\ 2012, \apj, 745, 73 

\bibitem[Nugent et al.(2011)]{2011Natur.480..344N} Nugent, P.~E., Sullivan, 
M., Cenko, S.~B., et al.\ 2011, \nat, 480, 344 

\bibitem[Pakmor et al.(2012)]{2012ApJ...747L..10P} Pakmor, R., Kromer, M., 
Taubenberger, S., et al.\ 2012, \apjl, 747, L10 

\bibitem[Pastorello et al.(2007)]{2007MNRAS.377.1531P} Pastorello, A., 
Mazzali, P.~A., Pignata, G., et al.\ 2007, \mnras, 377, 1531 

\bibitem[Patat et al.(2007)]{2007Sci...317..924P} Patat, F., Chandra, P., 
Chevalier, R., et al.\ 2007, Science, 317, 924 

\bibitem[Perlmutter et~al.(1999)]{perl99} Perlmutter, S. et~al. 1999, \apj, 517, 565 


\bibitem[Phillips et al.(1987)]{1987PASP...99..592P} Phillips, M.~M., 
Phillips, A.~C., Heathcote, S.~R., et al.\ 1987, \pasp, 99, 592 

\bibitem[Phillips(1993)]{1993ApJ...413L.105P} Phillips, M.~M.\ 1993, \apjl, 
413, L105 

\bibitem[Riess et~al.(1998)]{riess98} Riess, A.~G. et~al. 1998, \aj, 116, 1009

\bibitem[R{\"o}pke et al.(2012)]{2012ApJ...750L..19R} R{\"o}pke, F.~K., 
Kromer, M., Seitenzahl, I.~R., et al.\ 2012, \apjl, 750, L19 

\bibitem[Silverman et al.(2011)]{2011MNRAS.410..585S} Silverman, J.~M., 
Ganeshalingam, M., Li, W., et al.\ 2011, \mnras, 410, 585 

\bibitem[Simon et al.(2009)]{2009ApJ...702.1157S} Simon, J.~D., Gal-Yam, 
A., Gnat, O., et al.\ 2009, \apj, 702, 1157 

\bibitem[Sternberg et al.(2011)]{2011Sci...333..856S} Sternberg, A., 
Gal-Yam, A., Simon, J.~D., et al.\ 2011, Science, 333, 856 

\bibitem[Stritzinger et al.(2010)]{2010AJ....140.2036S} Stritzinger, M., 
Burns, C.~R., Phillips, M.~M., et al.\ 2010, \aj, 140, 2036 

\bibitem[Taubenberger et al.(2011)]{2011MNRAS.412.2735T} Taubenberger, S., 
Benetti, S., Childress, M., et al.\ 2011, \mnras, 412, 2735 

\bibitem[Vernet et 
al.(2011)]{2011A&A...536A.105V} Vernet, J., Dekker, H., D'Odorico, S., et al.\ 2011, \aap, 536, A105 

\bibitem[Wang et al.(2009)]{2009ApJ...697..380W} Wang, X., Li, W., 
Filippenko, A.~V., et al.\ 2009, \apj, 697, 380 

\bibitem[Yaron 
\& Gal-Yam(2012)]{2012arXiv1204.1891Y} Yaron, O., \& Gal-Yam, A.\ 2012, arXiv:1204.1891 

\end{thebibliography}
\end{document}